\documentclass[12pt]{article}

\usepackage{a4,graphicx,amsmath,amssymb}
\usepackage{cite}
\def\Bbb{\mathbb}
\def\BZ{\Bbb Z} \def\BR{\Bbb R}
\def\BC{\Bbb C} \def\BP{\Bbb P}

\catcode`@=11 \@addtoreset{equation}{section} \catcode`@=12

\begin{document}
\begin{titlepage}
\renewcommand{\thefootnote}{\fnsymbol{footnote}}
\noindent
{\tt IITM/PH/TH/2007/8}\hfill
{\tt arXiv:0707.4306v3}\\
{\tt IC/2007/059} \hfill August 2007\\
\begin{center}
\large{\bf  
Symplectic potentials and resolved Ricci-flat ACG metrics}
\end{center} 
\bigskip 
\begin{center}
Aswin K. Balasubramanian\footnote{Address after Aug. 20, 2007: 
Dept. of Physics, Univ. of Texas at Austin, Austin, TX 78712 USA,
\texttt{aswin@mail.utexas.edu}}, 
Suresh Govindarajan\footnote{\texttt{suresh@physics.iitm.ac.in}} \\
\textit{Department of Physics, Indian Institute of Technology Madras, Chennai 600036, INDIA.}\\
 and \\
Chethan N. Gowdigere\footnote{\texttt{cgowdige@ictp.it}}\\
\textit{Abdus Salam ICTP, Strada Costiera 11,
34014 Trieste, ITALY.}\\
\end{center}
\begin{abstract}
We pursue the symplectic description of toric K\"ahler manifolds. 
There exists a general local classification of metrics on toric 
K\"ahler manifolds equipped with Hamiltonian two-forms due to 
Apostolov, Calderbank and Gauduchon(ACG). We derive the 
symplectic potential for these metrics. Using a method due to 
Abreu, we relate the symplectic potential to the canonical 
potential written by Guillemin. This enables us to recover the 
moment polytope associated with metrics and we thus obtain global 
information about the metric. We illustrate these general 
considerations by focusing on six-dimensional Ricci-flat metrics 
and obtain Ricci-flat metrics associated with real cones over 
$L^{pqr}$ and $Y^{pq}$ manifolds. The metrics associated with 
cones over $Y^{pq}$ manifolds turn out to be partially resolved 
with two blow-up parameters taking special (non-zero)values. For a 
fixed $Y^{pq}$ manifold, we find explicit metrics for several 
inequivalent blow-ups parametrised by a natural number $k$ in the 
range $0<k<p$. We also show that all known examples of 
resolved metrics such as the resolved conifold and the resolution of
$\BC^3/\BZ_3$  also fit the ACG classification.
\end{abstract}
\end{titlepage}
\renewcommand{\thefootnote}{\arabic{footnote}}
\setcounter{footnote}{0}

\section{Introduction}

The natural target space for $(2,2)$ supersymmetric non-linear 
sigma models in two dimensions is a K\"ahler manifold, 
$X$\cite{Zumino:1979et}. For applications in string theory, one 
needs the non-linear sigma model to be conformally invariant. To 
leading order, conformality of the non-linear sigma model requires the 
K\"ahler manifold to be Ricci-flat\cite{Friedan:1980jm}. In 1993, 
Witten provided a simpler construction of such sigma models by 
introducing the gauged linear sigma 
model(GLSM)\cite{Witten:1993yc}. Among the many phases of the 
GLSM, he showed that there is a phase where one recovers the 
non-linear sigma model. A notable feature of this construction 
was a simple description of a necessary condition (i.e., 
$c_1(X)=0$) for the manifold to be Ricci-flat. Further, he showed 
that the GLSM naturally realises a symplectic quotient and that 
the induced metric in the NLSM limit was a natural generalisation 
of the Fubini-Study metric associated with complex projective 
spaces.

Around the same time, Guillemin 
carried out a systematic treatment of toric K\"ahler manifolds and
wrote a simple formula that  generalised the Fubini-Study metric for 
$\BC\BP^n$\cite{guillemin}. The only data that went into writing 
the metric was the moment polytope associated with a toric 
K\"ahler manifold. The moment polytope is a convex polytope 
defined by several inequalities of the form
\begin{equation}
\ell_a(P)>0\ ,\ a=1,2,\ldots
\end{equation}

Guillemin wrote the metric in symplectic coordinates rather than 
(the more commonly used) complex coordinates. The metric in 
symplectic coordinates is determined by a single function called 
the \textit{symplectic potential}\cite{AbreuReview}. The 
symplectic potential written by Guillemin is given by
\begin{equation}
\label{guillemin}
G_{\textrm{can}}(P)=\frac12 \sum_{a}\ell_a(P)\ \log
\, \ell_a(P)  \ .
\end{equation}
We will refer to this as the \textit{canonical} symplectic 
potential. For projective spaces, this metric reduces to the 
Fubini-Study metric. However, while the metric correctly captures 
the singularities in more general examples, it is not necessarily 
Einstein (or even extremal) as the Fubini-Study metric. It turns 
out that the metric given by the GLSM is identical to the one 
written out by Guillemin\footnote{This result may be obvious to 
some and non-obvious to others. However, the GLSM has a wider 
range of validity than the Guillemin formula. For instance, it is 
valid even for non-toric examples.}.

Abreu had a simple suggestion to obtain Einstein/extremal metrics 
from the canonical one\cite{Abreu:1997}. Adapting a method due to
Calabi in the complex context\cite{Calabi:1979}, Abreu 
modified the canonical symplectic potential by adding a `function' 
to it as follows:
\begin{equation}
\label{Abreu}
G(P) = G_{\textrm{can}}(P) + h(P)\ ,
\end{equation}
where $h(P)$ is \textit{non-singular} in the interior as well as 
the boundary of the polytope. We will refer to $h(P)$ as the 
\textit{Abreu} function in this paper. The Abreu function is 
determined by requiring that the new metric has the required 
property such as extremality. For instance, the differential 
equation for $h(P)$ is the analog of the Monge-Amp\`ere equation 
that appears when one imposes Ricci-flatness on the K\"ahler 
potential\cite{Donaldson}.  The function $h(P)$ has been 
determined in only a small number of 
examples\cite{Abreu:1997,Ray:1998pm,AKBSG}. However, there have 
been a recent attempt to obtain the function 
numerically\cite{Doran:2007zn}.

This paper focuses on a special sub-class of toric K\"ahler 
manifolds, those that admit a Hamiltonian 2-form. For K\"ahler 
manifolds that admit such a 2-form (and possibly non-toric), there exists 
a classification of these metrics due to 
Apostolov, Calderbank and Gauduchon(ACG)\cite{ACG1}. The main 
merit of such metrics is that it replaces a PDE in $m$ variables 
that one needs to solve to obtain the symplectic potential by an
ODE's in $m$ functions of one-variable in the best of situations.  
We obtain the symplectic potential for these metrics and find 
that it can be easily written in the form given in 
Eqn.(\ref{Abreu}). Then the associated polytope is easily 
recovered. We find that all known examples of resolved metrics
in six-dimensions admit a Hamiltonian 2-form and add a \textit{new}
infinite family of partially resolved spaces to the list of known examples.

Another application of these methods is in the context of the 
AdS-CFT correspondence which relates four-dimensional conformal 
field theories with type IIB string theory on $AdS_5\times X^5$, 
where $X^5$ is a compact five-dimensional Sasaki-Einstein 
manifold\cite{Maldacena:1997re}. Real cones over these spaces 
turn out to be non-compact Ricci-flat K\"ahler manifolds. Thus, 
our examples will focus on six-dimensional Ricci-flat toric 
K\"ahler manifolds which are allowed to have a conical 
singularity at the tip of the cone. Resolutions of these 
singularities correspond to non-conformal deformations of the 
conformal field theory and are also of independent 
interest\cite{Klebanov:1999tb}.

The paper is organised as follows. Section 2 is a review of the 
symplectic quotient as obtained from the gauged linear sigma 
model.  In section 3, we review the local classification of toric 
K\"ahler metrics admitting a Hamiltonian 2-form due to Apostolov, 
Calderbank and Gauduchon. We then discuss the conditions 
under which their metrics are Einstein and Ricci-flat. In section 
4, we construct the symplectic potential for all their metrics. 
We then carry out a global analysis of the ACG metrics and 
discuss how one recovers the precise singularity structure by 
writing the symplectic potential as the sum of the canonical symplectic potential
and the Abreu function. Sections 5 and 6 make use of the results of section 4 to 
generate examples of unresolved and resolved metrics 
respectively. While the results in section 5 are not new, the 
methods used are new and have independent merit. In section 6, we 
obtain a new infinite family of metrics corresponding to partially 
resolved metrics on cones over $Y^{pq}$. We conclude in section 7 
with a brief discussion on our results.

\noindent \textbf{Note added:} While this paper was being readied 
for publication, a paper by Martelli and Sparks 
appeared\cite{Martelli:2007pv}. This paper also deals with ACG 
metrics and resolved metrics. There is some overlap with this 
work though the methods differ. The authors also mention a 
forthcoming paper which discusses the partial resolutions of 
cones over $Y^{pq}$ spaces. This also may have some overlap with 
section 6 of this paper.

\section{The symplectic quotient in the GLSM}

A large family of K\"ahler manifolds are obtained by means of the K\"ahler
quotient. The construction proceeds as follows\cite{Morrison:1994fr}:
\begin{equation}
X^{(2m)} = \frac{\BC^n -F_\Delta}{(\BC^*)^d}\ , \quad m=n-d\ .
\end{equation}
The various $\BC^*$ actions are specified by the \textit{charge 
vectors} ${Q_a}^\alpha$(which we sometimes write as a $n \times 
d$ matrix $\mathcal{Q}$):
\begin{equation}
\phi_a \longrightarrow \lambda^{{Q_a}^\alpha}\  \phi_a \quad,\  
a=1,\ldots,n\textrm{ and }
\alpha=1,\ldots,d
\end{equation}
$F_\Delta$ corresponds to the set of fixed points under the 
$\BC^*$ actions. For instance, $\BC\BP^{n-1}$ is obtained by the 
K\"ahler quotient with one $\BC^*$ action with charge vector 
$Q=(1,1,\ldots,1)^T$ and $F_\Delta =\{0\}$.

Writing $\BC^*=\BR_+\times S^1$, the $\BC^*$ quotient can be 
carried out as a two-step process. First, carry out $\BR^+$ 
quotient and then the $S^1$ action. This is called the {\em 
symplectic quotient} and this is the way the GLSM naturally 
realises the quotient\cite{Witten:1993yc}.

The symplectic quotient is implemented in the GLSM as follows. 
The GLSM has $(2,2)$ supersymmetry and the field content consists 
of $n$ chiral superfields, $\Phi_a$ ($a=1,\ldots,n$) and $d$ 
Abelian vector multiplets $V_\alpha$ ($\alpha=1,\ldots,d$). 
(Please see \cite{Witten:1993yc} for more details.) The charges 
of the chiral fields under the $d$ gauge fields is given by $d$ 
charge vectors, ${Q_a}^\alpha$ , $\alpha=1,\ldots,d$. The 
parameters of the GLSM are the gauge coupling constant $e^2$ (we 
take all the $d$ couplings to be identical for simplicity). Each 
abelian multiplet admits a Fayet-Iliopoulos (FI) term which is 
represented by a complex coupling, $\tau_\alpha\equiv r_\alpha + 
i (\theta_\alpha/2\pi)$. We will refer to $r_\alpha$ as FI 
parameters or blow-up parameters.

In the GLSM, the $(\BR_+)^d$ quotient is imposed by the D-term 
constraints\footnote{In the strong coupling limit(s), typically 
of the form $e^2 r_\alpha\rightarrow \pm \infty$, the fields in 
the vector multiplets become Lagrange multipliers imposing 
various constraints (explicitly given, for instance, in 
\cite{Govindarajan:2000ef}).}:
\begin{equation}
\sum_{a=1}^n {Q_a}^\alpha |\phi_a|^2 = r_\alpha\ .
\end{equation}
and the $(S^1)^d\sim U(1)^d$ action is taken care of by the 
gauging in the GLSM. Not all values of $|\phi_i|^2$ can satisfy the D-term 
conditions. The set of allowed values of $|\phi_i|^2$ are best 
represented by the interior points of a convex polytope, the {\bf 
moment polytope}.  Writing the complex field 
$\phi_a=\sqrt{\ell_a} \exp{i\varphi_a}$ in polar coordinates, the 
D-term conditions can be written as
\begin{equation}
\sum_{a=1}^n {Q_a}^\alpha \ell_a  =r_\alpha\  ,
\end{equation}
These linear conditions can be solved for in terms of $m=n-d$ 
independent variables that we will call $P_i$ ($i=1,\ldots,m$). 
We can then rewrite the $\ell_a$ as implicit functions of the 
$P_i$, $\ell_a(P)$. The moment polytope is then given by the 
conditions
\begin{equation}
\ell_a(P) > 0\ .
\end{equation}
The K\"ahler two-form on the toric manifold $X^{(2m)}$ is 
$$
\omega =  \sum_{i=1}^m d P_i \wedge d t_i\ .
$$
where $t_i$ are the angles that remain after the $U(1)^d$ gauge 
degrees are removed. The metric on $X^{2m}$ is determined by a 
single function, $G(P)$, called the symplectic 
potential\footnote{The symplectic potential is the analogue of 
the K\"ahler potential appearing in complex coordinates. The two 
are related by a Legendre transformation\cite{AbreuReview}.}
\begin{equation}\label{symplecticmetric}
ds^2 = G_{ij} dP_idP_j + G^{ij} dt_idt_j\ ,
\end{equation}
where $G_{ij}= \partial^2 G/\partial P_i\partial P_j$ and 
$G^{ij}$ is the matrix inverse of $G_{ij}$. The metric induced by 
the symplectic quotient is the canonical symplectic potential 
$G_{\textrm{can}}$ given in Eqn. (\ref{guillemin}).

There is a theorem due to Delzant that states that one can 
recover a compact toric symplectic K\"ahler manifold from its 
polytope provided it satisfies certain conditions such as 
convexity, simplicity etc\cite{Delzant:1988}. Such polytopes have 
been called Delzant polytopes. An extension of Delzant's theorem 
to include toric symplectic orbifolds leads to polytopes with a 
positive integer attached to each facet\cite{Lerman:1994}. 
Weighted projective spaces have polytopes of this kind. The 
formula of Guillemin, given in Eqn. (\ref{guillemin}) though 
originally written out only for Delzant polytopes is valid for 
toric symplectic orbifolds as well\cite{Calderbank:2003}. The 
$P_i$ are thus coordinates on the polytope and the toric manifold 
is a $U(1)^m$ fibration with base, the polytope. The boundaries 
of the polytope correspond to points where the fibration 
degenerates\cite{Leung:1997tw}.

The usual toric data associated with toric manifold $X^{2m}$ is 
specified given by a set of $n$ vectors in $\BR^m$, written out 
as a $m\times n$ matrix $\mathcal{V}$.  These vectors are 
obtained from the charge vectors ${Q_a}^\alpha$ by solving
\begin{equation}
\label{nullspace}
\mathcal{V} \cdot \mathcal{Q} =0\ .
\end{equation}
Thus, while the charge vectors $\mathcal{Q}$ appear naturally in 
the GLSM, the toric description is given in terms of 
$\mathcal{V}$. In our examples, we will go back and forth between 
the two objects.

\subsection{Six-dimensional Ricci-flat manifolds}

A simple class of six-dimensional manifolds are obtained by 
considering the symplectic quotient involving the D-term given by 
the charge vector $Q=(p_1,p_2,p_3,p_4)^T$:
\begin{equation}
p_1 \ell_1 + p_2 \ell_2 + p_3\ell_3 + p_4 \ell_4= r\ ,
\end{equation}
where the $p_a$ are taken to be integers. This corresponds to the 
symplectic quotient $\BC^4/\BC^*$. A necessary condition for the 
manifold to admit a Ricci-flat metric is the condition 
$p_1+p_2+p_3+p_4=0$. There are two inequivalent classes of these 
four integers: (i) $p_1,p_2>0$ and $p_3,p_4<0$ and (ii) 
$p_1,p_2,p_3>0$ and $p_4<0$. All other possibilities can be 
obtained by suitably relabelling the $p_a$. The first choice 
leads to the conifold and its generalisations corresponding to 
real cones over $L^{pqr}$ spaces\cite{Cvetic:2005ft} and the 
second choice leads to orbifolds of the form $\BC^3/\BZ_{N}$ with 
$N=p_4$.

A blow-up is implemented in the GLSM by adding a new chiral 
superfield along with an additional abelian vector superfield. 
This adds a new D-term and leads to the symplectic quotient 
$\BC^5/(\BC^*)^2$. The FI parameter of the new D-term 
determines the size of the 
blown-up manifolds. In this paper, we will consider this 
situation as well.

\section{The ACG metrics.}

In this section, we will summarize the results from the paper 
\cite{ACG1} that are relevant for our purposes. The paper 
\cite{ACG1} concerns Hamiltonian 2-forms and a local 
classification of K\"ahler metrics that admit such 2-forms.
On a K\"ahler manifold $X$  of real dimension $2m$ with metric $g_{ij}$,
complex structure ${J^j}_i$ and K\"ahler two-form $\omega_{ij}=g_{jk}{J^k}_i$,
a \textit{Hamiltonian 2-form}, $\phi_{ij}$, is a $(1,1)$ form 
satisfying the equations
\begin{eqnarray}
\nabla_k \phi_{[ij]}  
+ \frac12 \left[  \partial_{[i} \textrm{Tr}\phi ~\omega_{j]k} +
\partial_l \textrm{Tr}
\phi ~J^l_{~[i}\, g_{j]\,k} \right] = 0\ .
\end{eqnarray}
where $\textrm{Tr}(\phi)=\omega^{kl}\phi_{kl}$ and $\nabla_k$ is the
covariant derivative with respect to the
Levi-Civita connection.

The notion of a Hamiltonian 2-form and the special 
properties of this object first appeared in \cite{ACG2}, where 
the authors were investigating a special class of four-dimensional 
K\"ahler metrics. It turned out that the Ricci form of this class 
of K\"ahler metrics was a Hamiltonian 2-form. The nomenclature 
``Hamiltonian" alludes\footnote{For the various equivalent and 
more precise definitions, see \cite{ACG1,ACG2}} to the fact that 
two scalars constructed out of the two-form: the trace, $s$, and 
the Pfaffian, $p$, in the four-dimensional context, are 
Hamiltonian functions for (Hamiltonian) Killing vector fields of 
the K\"ahler metric. The scalar functions also arise as the 
co-efficients of the characteristic polynomial of the $2 \times 2$ 
hermitian matrix, $(\phi-t\, \omega)$, constructed out of the Hamiltonian 2-form and 
the K\"ahler form:
\begin{equation}\label{1}
p(t) := t^2 - P_1 \, t + P_2.
\end{equation}
More importantly, the roots of this polynomial, call them, $\xi$ 
and $\eta$, so that
\begin{equation}\label{2}
P_1 = \xi + \eta, \qquad P_2 = \xi \, \eta,
\end{equation}
provide coordinates in which it becomes possible to classify a 
sub-class of toric K\"ahler metrics known as \emph{orthotoric} 
metrics in terms of two polynomials of one variable, one of $\xi$ 
and the other of $\eta$. An orthotoric metric is one with 
$g^{\xi\,\eta} = 0$ and the most general four-dimensional 
orthotoric metric is \cite{ACG2}
\begin{equation}\label{3}
ds_{\textrm{OT2}}^2 = (\xi - \eta) \left( \frac{d\xi^2}{f(\xi)} 
- \frac{d\eta^2}{g(\eta)} \right)  
+ \frac{1}{\xi - \eta} \left( f(\xi) (dt + \eta \,dz)^2 - g(\eta) ( dt 
+ \xi \,dz)^2 \right).
\end{equation}

In \cite{ACG1}, ACG worked in arbitrary dimensions and classified 
all $2\,m$-dimensional K\"ahler metrics which admit Hamiltonian 
2-forms. 
The existence of a Hamiltonian 2-form leads to
the existence of $m$ (Hamiltonian) Killing vector fields that commute. 
The co-efficients of the ``momentum polynomial,"  
$p(t)=\textrm{det}(\phi -t\, \omega)$, are the Hamiltonian functions
for the Killing vector fields. Further the roots of the momentum polynomial
provide special coordinates which permit an explicit 
classification of the metric. In general situations, 
the $m$ Killing vector fields may not all be linearly independent. 
A Hamiltonian two-form of order  $l\leq m$
leads to $l$ linearly independent Killing vector fields.
Thus  some (i.e., $(m-l)$) of the roots of the 
momentum polynomial are constants and hence can't provide for 
coordinates.  
  
ACG have shown
that the existence of a Hamiltonian two-form of order $l$ implies that \\
(i) the K\"ahler metric on $X$ can locally be written as a fibration(using a
construction due to Pedersen and Poon\cite{Pedersen:1991}), with a
$2l$-dimensional toric fibre over a $(2m-2l)$ dimensional base, \\
(ii) the K\"ahler
structure of the manifold, i.e., $(g,J,\omega)$ are completely specified
by $l$ functions of one-variable and the K\"ahler structure of the base. 

When $l=m$, the
manifold is necessarily toric though not all toric manifolds admit
a Hamiltonian 2-form of order $l=m$.
Thus such manifolds are called
\textit{orthotoric} reflecting the extra structure. The results of \cite{ACG2}, 
\eqref{3} is the special case, $m = 2, ~ l = 2$. The other extreme,
$l=0$ is the situation with no Killing vector fields. Thus, the
results of  ACG provides a nice classification of K\"ahler manifolds
that takes one from manifolds with no symmetries to orthotoric K\"ahler
manifolds.

In this paper since we are interested mainly in metrics on
six-dimensional manifolds, we 
will focus on the case $m = 3$, when the possible values for $l = 
1, 2, 3$. In \cite{ACG1}, the term orthotoric is used for the $l 
= m $ case and we shall do the same. We will add a subscript 
`OT$m$' to indicate the $2m$-dimensional orthotoric metric.  In all 
other situations, we will indicate the values of $m$ and $l$ in 
the subscript. The momentum polynomial has no constant roots
\begin{equation}\label{4}
p(t) = (t - \xi)\, (t - \eta) \, (t- \chi)
\end{equation}
and the Hamiltonian functions for the Killing vector fields 
$\frac{\partial}{\partial t_1}, \frac{\partial}{\partial t_2}, 
\frac{\partial}{\partial t_3}$ are
\begin{equation}
P_1 = \xi + \eta + \chi, \qquad P_2 = \xi\,\eta + \eta \, 
\chi + \chi \, \xi, \qquad P_3 = \xi\,\eta\,\chi
\end{equation}
The most general orthotoric metric admitting a Hamiltonian 2-form 
is then given in terms of three polynomials of one variable:
\begin{align}
ds_{\textrm{OT3}}^2 =& - \Delta\left[ \frac{d\xi^2}{ (\eta - \chi) f(\xi)} 
+ \frac{d\eta^2}{(\chi - \xi) g(\eta)} + \frac{d\chi^2}{(\xi 
- \eta) h(\chi)} \right] \nonumber \\ 
&- \frac{1}{\Delta } \Big[ (\eta - \chi) \, f(\xi)\,  
(dt_1 + (\eta + \chi) dt_2 + \eta\,\chi \, dt_3)^2  \nonumber \\
 &\hspace{1cm} +  (\chi - \xi) \, g(\eta)\,  
(dt_1 + (\chi + \xi) dt_2 + \chi\,\xi \, dt_3)^2 \nonumber \\
  &\hspace{1cm}+  (\xi - \eta) \, h(\chi)\,  
(dt_1 + (\xi + \eta) dt_2 + \xi\,\eta \, dt_3)^2 \Big]\ . 
  \end{align}
where $\Delta=(\xi - \eta) \, (\eta - \chi ) \, ( \chi - \xi )$. The K\"ahler 
form, the Hamiltonian 2-form
and the scalar curvature ($R$) for the $m=3$, $l = 3$ ACG metrics are given by 
\begin{eqnarray}
\omega_{\textrm{OT3}}
 &= & dP_1 \wedge dt_1 + dP_2 \wedge dt_2 +dP_3 \wedge dt_3\,, \\
\phi_{\textrm{OT3}} &=& 
\big[P_1dP_1-dP_{2}\big]\wedge dt_1
+\big[P_2dP_1-dP_{3}\big]\wedge dt_2
+P_3dP_1\wedge dt_3\ , \\
R_{\textrm{OT3}} &=& 
-\frac{f''(\xi)}{(\xi-\eta)(\xi-\chi)} 
-\frac{g''(\eta)}{(\eta-\xi)(\eta-\chi)}
-\frac{h''(\chi)}{(\chi-\eta)(\chi-\xi)}\, .
\end{eqnarray}

We will also need the $m = 3$, $l = 2$ case, when the momentum polynomial is 
\begin{equation}\label{5}
p(t) = (t - a) \, (t - \xi) \, ( t - \eta), \qquad a = \textrm{constant}.
\end{equation}
We then have only two Hamiltonian functions for the Killing 
vector fields $\frac{\partial}{\partial t_1}, 
\frac{\partial}{\partial t_2}$ viz. $P_1 = \xi + \eta, ~~P_2 = 
\xi \, \eta$. Thus the roots of the momentum polynomial will 
provide two of the coordinates for the local description. We will 
refer to these K\"ahler metrics admitting Hamiltonian 2-forms as 
the \emph{$m=3$, $l = 2$ ACG} metrics. The most general $m=3$, $l = 2$ ACG 
metric is then given by \cite{ACG1}
\begin{align}\label{6}
ds_{m=3,l=2}^2 =& ( a - \xi )\,(a - \eta) ds^2_a + (\xi 
- \eta) \left[ \frac{\xi - a}{f(\xi)} \, d\xi^2 
- \frac{\eta -a }{g(\eta)}\, d\eta^2 \right]\nonumber  \\
&+ \frac{1}{\xi - \eta} \left[ \frac{f(\xi)}{\xi - a} 
(\theta_1 + \eta \,\theta_2)^2 - \frac{g(\eta)}{\eta - a} 
( \theta_1 + \xi \,\theta_2)^2 \right],
\end{align}
where $ds^2_a$ is a K\"ahler metric on a two-dimensional manifold
with a K\"ahler form 
$\omega_a$, $\theta_1$ and $\theta_2$ are one-forms which satisfy 
the following conditions,
\begin{equation}\label{7}
d\, \theta_1 = - a~ \omega_a, \qquad d\,\theta_2 =  ~\omega_a\,. 
\end{equation}
The K\"ahler form, Hamiltonian 2-form and the scalar curvature for the $m=3$, $l = 2$ ACG metrics are
given by
\begin{eqnarray}
\omega_{m=3,l=2} &=& ( a - \xi )\,(a - \eta) \omega_a + 
d\,(\xi + \eta) \wedge \theta_1 + d\,(\xi\,\eta) \wedge \theta_2\,, \\
\phi_{m=3,l=2} &=& \!\!a ( a - \xi )\,(a - \eta) \omega_a + 
\big[P_1dP_1-dP_{2}\big]\wedge d\theta_1
+P_2dP_1\wedge d\theta_2
\\
R_{m=3,l=2} &=& \frac{R\big(ds^2_a\big)}{(a-\xi)(a-\eta)}-\frac{f''(\xi)}{(\xi-\eta)(\xi-a)} 
-\frac{g''(\eta)}{(\eta-\xi)(\eta-a)}\ ,
\end{eqnarray}
where $R(ds^2_a)$ is the scalar curvature of the metric $ds^2_a$.

We will also need the $m = 3, l =1$ case, when there are 
\textit{two} possibilities for the momentum polynomial:
\begin{eqnarray}\label{p1}
p_1(t) &=& (t - a)^2 \, (t - \xi), \qquad a = \textrm{constant},\\ \label{8}
p_2(t) &=& (t - a) \, (t - b) \, (t - \xi), \qquad a,b = \textrm{constants }, 
\end{eqnarray}
with $a< b$. We then have only one Hamiltonian function for the 
Killing vector field $\frac{\partial}{\partial t_1}$ viz. $P_1=\chi$. 
We will refer to these K\"ahler metrics admitting Hamiltonian 
2-forms as the \emph{$m=3$, $l = 1$ ACG} metrics . The most general $m=3$, $l = 
1$ ACG metric is of either of two types depending on the momentum 
polynomial. For \eqref{p1}, the most general $m=3$, $l =1$ ACG metric is 
given by \cite{ACG1},
\begin{equation}\label{m3l1a}
\Big[ds^2_{m=3,l=1}\Big]_{p_1(t)} = (a - \chi)\, ds^2_a 
+ \frac{(\chi - a)^2}{h(\chi)} d\chi^2 + \frac{h(\chi)}{(\chi - a)^2} \, 
\theta_1^2,
\end{equation}
where $ds^2_a$ is a K\"ahler metric for the four-dimensional base
with K\"ahler form 
$\omega_a$ and $\theta_1$ is a one-form which satisfies
\begin{equation}\label{10}
d\theta_1 = - \, \omega_a.
\end{equation}
The K\"ahler form, Hamiltonian 2-form and scalar curvature for the above $m=3$, $l =1$ ACG metrics are 
\begin{eqnarray}
\big[\omega_{m=3,l=1}\big]_{p_1(t)} &=& (a - \chi) \, \omega_a + d\chi \wedge \theta_1 \ ,\\
\big[\phi_{m=3,l=1}\big]_{p_1(t)} &=& a(a - \chi) \, \omega_a + \chi d\chi \wedge \theta_1 \ ,\\
\big[R_{m=3,l=1}\big]_{p_1(t)}&=& \frac{R\big(ds^2_a\big)}{(a-\chi)}
 -\frac{h''(\chi)}{(\chi-a)^2}\,,
\end{eqnarray}
where $R(ds^2_a)$ is the scalar curvature of the metric $ds^2_a$.

For the momentum polynomial \eqref{8}, the most general $m=3$, $l = 1$ 
ACG metric is given by \cite{ACG1},
\begin{equation}\label{m3l1b}
\Big[ds^2_{m=3,l=1}\Big]_{p_2(t)} = (a - \chi) \, ds^2_a + (b 
- \chi)\,  ds^2_b + \tfrac{(\chi - a) (\chi - b)}{h(\chi)} \, 
d\chi^2 + \tfrac{h(\chi)}{(\chi - a) (\chi - b)} \, \theta_1^2,
\end{equation}
where $ds^2_a$ and $ds^2_b$ are two K\"ahler metrics with 
K\"ahler forms $\omega_a$ and $\omega_b$ with the one-form 
$\theta_1$ satisfying
\begin{equation}\label{12}
d\theta_1 = - \, \omega_a - \, \omega_b.
\end{equation}
The K\"ahler form, Hamiltonian 2-form and the scalar curvature  for the above $m=3$, $l =1$ ACG metrics are given by 
\begin{eqnarray}
 \big[\omega_{m=3,l=1}\big]_{p_2(t)} &=& (a - \chi)\, \omega_a + ( b - \chi)\, 
\omega_b + d\chi \wedge \theta_1 \ ,\\
 \big[\phi_{m=3,l=1}\big]_{p_2(t)} &=& a(a - \chi)\, \omega_a + b( b - \chi)\, 
\omega_b + \chi d\chi \wedge \theta_1 \ ,\\
 \big[R_{m=3,l=1}\big]_{p_2(t)}&=& \frac{R\big(ds^2_a\big)}{(a-\chi)}
  + \frac{R\big(ds^2_b\big)}{(b-\chi)}
 -\frac{h''(\chi)}{(\chi-a)(\chi-b)}\,.
\end{eqnarray} 

\subsection{Imposing extra conditions on the ACG metrics}

In the sequel, we will find examples for resolutions of metric 
cones in the $m=3$, $l = 2$ ACG case \eqref{6} and the $m=3$, $l 
=1$ ACG cases \eqref{m3l1a} and \eqref{m3l1b}. We will therefore 
gather some more facts about these cases, mainly 
restrictions imposed on the polynomials $f(\xi)$, $g(\eta)$ and 
$h(\chi)$ by conditions such as Ricci-flatness, Einstein, etc.

\subsubsection*{Extremality}

A K\"ahler metric is said to be \textit{extremal} when the 
scalar curvature is the Hamiltonian function for 
a Killing vector field. In our examples, this implies that
the scalar curvature is an affine function of the
Hamiltonian function, $P_1$. For instance for the OT3 metric, this occurs when
$f''(x)=g''(x)=h''(x)$ and the functions $f/g/h$ have degree four. For all other
metrics, additional conditions arise and are described below.

If further, one requires a slightly stronger condition (as ACG do) than the one
required by extremality, i.e.,  $f'(x)=g'(x)=h'(x)$, then the normalized
Ricci form, 
$$
\tilde{\rho}\equiv \rho-\frac{R\omega}{8}\ ,
$$ 
is a Hamiltonian 2-form and can be written as a linear combination of 
the Hamiltonian 2-form, $\phi$  and the K\"ahler form.  

\subsubsection*{The Einstein condition}

The ACG metrics are Einstein metrics when the following three 
conditions are satisfied: 
\begin{itemize} \item[\textbf{(i)}] 
$f'(x) = g'(x)=h'(x)$ and furthermore they should factorize in 
the following way: 
\begin{eqnarray} 
m=3, l=3,\quad f'(x)&=& b_{-1} x^4 
+ b_0 x^3 + b_1 x^2 +b_2 x + b_3 \nonumber \\ 
m=3, l = 2, \quad f'(x) 
&=& (x - a) (b_{-1} \, x^3 + b_0 \, x^2 + b_1 \, x + b_2)  
\nonumber \\ 
m=3, l = 1~ \textrm{with}~ p_1(t), \quad h'(x) &=& (x - 
a)^2 \, (b_{-1} \, x^2 + b_0 \, x + b_1 \, )  \\ 
m=3, l = 1~ 
\textrm{with}~ p_2(t), \quad h'(x) &=& (x - a)\,(x-b) \, (b_{-1} 
\, x^2 + b_0 \, x + b_1 )  \nonumber 
\end{eqnarray} 
for some constants $b_i$. 
\item[\textbf{(ii)}] the smaller K\"ahler 
metrics, $ds^2_a$, $ds^2_b$ should be K\"ahler-Einstein with their 
scalar curvatures satisfying the following relations:
\begin{eqnarray}
m=3, l = 2, \qquad -R(ds^2_a) &=&  b_{-1} \, a^3 + b_0 \, a^2 + b_1 \, a + b_2 
\nonumber \\
m=3, l = 1~ \textrm{with }~ p_1(t), \qquad -\frac{R(ds^2_a) }{2} &=&  
b_{-1} \, a^2 + b_0 \, a + b_1 \nonumber \\
m=3, l = 1~ \textrm{with }~ p_2(t), \qquad -R(ds^2_a) &=& b_{-1} 
\, a^2 + b_0 \, a + b_1 ~\textrm{and} \nonumber \\ 
-R(ds^2_b) &=& b_{-1} \, b^2 + b_0 \, b + b_1\ .
\end{eqnarray} 
\item[\textbf{(iii)}] The Ricci-form is then given by
\begin{equation}
\rho = -\frac12 \big[b_{-1}(\phi + P_1 \omega)+b_0\omega\big]\ .
\end{equation}
which clearly leads to an Einstein metric when
\begin{equation}b_{-1} = 0\ .\end{equation}
\end{itemize}
The scalar curvature for these Einstein manifolds then is
equal to $-3b_0$.

\subsubsection*{The Ricci-flatness condition}

For the ACG metrics to be Ricci-flat as well, one needs
 \begin{equation}
b_0 = 0\ .
\end{equation}
As we will be interested in Ricci-flat metrics, we note that we 
will end up with the functions $f/g/h$ being cubic polynomials. 
This is all we need for orthotoric metrics. For the ACG metrics 
with $l<m=3$, it is useful to explicitly write out the conditions 
that are imposed on the smaller K\"ahler metrics.
\begin{enumerate}
\item When $m=3$, $l=2$, One needs $R(ds_a^2)=-b_1 a - b_2$.
\item When  $m=3$, $l=1$ with  polynomial  $p_1(t)$, one needs 
$R(ds_a^2)=-2b_1$.
\item When  $m=3$, $l=1$ with polynomial  $p_2(t)$, one needs 
$R(ds_a^2)=R(ds_b^2)=-b_1$.
\end{enumerate}

\section{Symplectic potentials for the ACG metrics}

In this section, we compute the symplectic potentials for the ACG 
metrics\footnote{The symplectic potential for the $Y^{pq}$ and 
$L^{pqr}$ metrics has been obtained in ref. \cite{Oota:2005mr}.}. 
We first start with the orthotoric case with $l=m=2$ since this 
is the first non-trivial orthotoric metric. The symplectic 
potential for higher-dimensional orthotoric metrics and other ACG 
metrics follows from this case.

\subsection{The $m=2$ orthotoric symplectic potential}

The coordinate transformation that gives $(\xi,\eta)$ as a 
function of $(P_1,P_2)$ is obtained by identifying $(\xi,\eta)$ 
with the roots of the quadratic equation
\begin{equation}
\lambda^2 - P_1 \lambda + P_2 =0\ .
\end{equation}

The $m=2$ orthotoric metric in the coordinates $(P_1,P_2)$ is given by
\begin{equation}
ds_{\textrm{OT2}}^2 = g_{ij} dP_i dP_j + g^{ij} d t_i d t_j
\end{equation}
where
$$
g_{ij} =\frac{1}{\xi-\eta}\begin{pmatrix} 
\big[\frac{\xi^2}{f(\xi)} -  \frac{\eta^2}{g(\eta)} \big] &
-\big[\frac{\xi}{f(\xi)} -  \frac{\eta}{g(\eta)} \big] \\
-\big[\frac{\xi}{f(\xi)} -  \frac{\eta}{g(\eta)} \big] &
\big[\frac{1}{f(\xi)} -  \frac{1}{g(\eta)} \big]
\end{pmatrix}
$$
We can check that this metric can indeed be written as the Hessian of a
symplectic potential. The integrability condition is given by
$$
\partial_i \big(g_{jk}\big)= \partial_j \big(g_{ik}\big)\ ,
$$
which holds in our case. 

We will write the symplectic potential as an explicit function of 
$(\xi,\eta)$ and hence as an implicit function of $(P_1,P_2)$. Let 
$G(\xi,\eta)$ be the symplectic potential for the above metric. 
Then, one has
\begin{align}
&g_{ij} = \partial_i \partial_j G(\xi,\eta) \\
&=\Big[
\frac{\partial^2 G}{\partial\xi^2} \partial_i \xi \partial_j \xi +
\frac{\partial^2 G}{\partial\xi\partial\eta} 
(\partial_i \xi \partial_j \eta +
\partial_j \xi \partial_i \eta) +
\frac{\partial^2 G}{\partial\eta^2} \partial_i \eta \partial_j \eta
\Big] + 
\Big[\frac{\partial G}{\partial\xi} \partial_i  \partial_j \xi +
\frac{\partial G}{\partial\eta} \partial_i  \partial_j \eta\Big]\nonumber \ .
\end{align}

The easiest of the three partial differential equations turns out to be
the one for $g_{22}$ which reads
\begin{equation}
\Big[ \bigg(\frac{\partial}{\partial\xi}-\frac{\partial}{\partial\eta}\bigg)^2 
-\frac{2}{\xi-\eta} \Big(\frac{\partial}{\partial\xi}
-\frac{\partial}{\partial\eta}\Big)\Big] G
= (\xi -\eta) \Big(\frac1{f(\xi)} -\frac1{g(\eta)}\Big)\ .
\end{equation}
Let us assume that $f$ and $g$ are polynomial functions with distinct roots
$(\xi_1,\ldots,\xi_N)$ and 
$(\eta_1,\ldots,\eta_{\tilde{N}})$ respectively. 
We will need the inverses of $f$ and $g$, which we write as ($f_0$ and $g_0$ 
are constants 
that turn out to be proportional to the scalar curvature. We deliberately include a
minus sign so that the constants are positive in our examples.)
\begin{eqnarray}
\frac{1}{f(\xi)} &=& \frac1{-f_0\prod_{a=1}^N (\xi - \xi_a)} =
\sum_{a=1}^N \frac{A_a}{(\xi - \xi_a)}\  , \\
\frac{1}{g(\eta)} &=& \frac1{-g_0\prod_{a=1}^{\tilde{N}} (\eta - \eta_a)} =
\sum_{a=1}^{\tilde{N}} \frac{B_a}{(\eta - \eta_a)}\ , 
\end{eqnarray}
where 
\begin{equation}
\label{ABdefs}
A_a\equiv -f_0^{-1}\prod_{b\neq a}(\xi_a -\xi_b)^{-1}\ , \textrm{ and  }
B_a\equiv -g_0^{-1}\prod_{b\neq a}(\eta_a -\eta_b)^{-1}\ .
\end{equation}

Since the partial differential equations are linear, we can use 
superposition. So, all we really need to do is to solve for the 
simple case when $f=(\xi-\xi_1)$ and dropping the term involving 
$g(\eta)$. It turns out that this is solved by the function 
$\mathcal{S}$ which we define as follows:
\begin{equation}
\label{simplepart}
\mathcal{S}(\xi,\eta,a) = \frac12 (\xi-\eta)(\xi-a) - 
(\eta -a)(\xi-a)\log (\xi-a)\ .
\end{equation}
This is the solution to the differential equation
\begin{equation}
\Big[ \Big(\frac{\partial}{\partial\xi}-\frac{\partial}{\partial\eta}\Big)^2
-\frac{2}{\xi-\eta} \bigg(\frac{\partial}{\partial\xi}
-\frac{\partial}{\partial\eta}\bigg)\Big] \mathcal{S}(\xi,\eta,\xi_1) 
=\frac{\xi -\eta}{\xi-\xi_1} \ .
\end{equation}
One can verify that the other partial differential equations for $g_{11}$ and
$g_{12}$ are also satisfied. Thus, we can now write the symplectic potential
explicitly as follows:
\begin{equation}
G(\xi,\eta) = 
\sum_{a=1}^{N} A_a\ \mathcal{S}(\xi,\eta,\xi_a)
+\sum_{a=1}^{\tilde{N}} B_a\ \mathcal{S}(\eta,\xi,\eta_a)\ .
\end{equation}
When $N$ and $\tilde{N}$ are both greater than two, a slightly simpler
form follows on using the identities $\sum_a A_a = \sum_a A_a \xi_a =0$
and a similar one for the $B$'s. In this case, we can write,
\begin{equation}
\label{sympotOT2}
G_{\textrm{OT2}}(\xi,\eta) = -\sum_{a=1}^N A_a\ (\eta-\xi_a)(\xi-\xi_a)\log 
(\xi-\xi_a)
 -\sum_{a=1}^{\tilde{N}} B_a\ (\xi-\eta_a)(\eta-\eta_a)\log (\eta-\eta_a)
\end{equation}
where $\xi_a$(resp. $\eta_a$) are the \textit{distinct} roots of 
$f(\xi)$(resp. $g(\eta)$). Note that the coefficient of each of 
the logarithms can be rewritten in terms of linear functions of 
$(P_1,P_2)$. For instance,
\begin{equation}
(\xi-\xi_1)(\eta-\xi_1)= P_2 -P_1 \xi_1 + \xi_1^2=p(\xi_1)\ ,
\end{equation}
where the last term is the momentum polynomial for this case, 
i.e., $p(t)=(t-\xi)(t-\eta)$. This observation enables us to 
analyse global properties of the orthotoric as well as ACG 
metrics.

We conclude this discussion with a comment on the symplectic 
potential for the other cases such as the $m=3$ orthotoric 
metric. This involves adding a piece corresponding to the roots 
of the third function, $h(\chi)$ and pre-multiplying the argument 
of all logarithms so that they can be written as the linear function of 
$(P_1,P_2,P_3)$ given by the momentum polynomial, $p(t)$.

\subsection{The $m=3$, $l=2$ symplectic potential}

In order to be more concrete, we choose the two-dimensional 
metric $ds^2_a$ to be the Fubini-Study metric for a $\BC\BP^1$ 
with symplectic potential
\begin{equation}
G_{FS}(x)=\frac12 (1-x)\ \log(1-x) + \frac12 (1+x) \log(1+x) \ .
\end{equation}
The scalar curvature for the above metric is 2.
The natural choice for the symplectic coordinates which follows 
from the K\"ahler form is
\begin{equation}
\label{Ypqcoords}
P_1=(\xi+\eta)\ , \  P_2=\xi\eta \textrm{ and } P_3=(a-\xi)(a-\eta)x\ .
\end{equation} 

A calculation similar to the one used to derive the symplectic 
potential for the $m=2$ orthotoric case leads to the following 
symplectic potential for the $m=3$, $l=2$ ACG metric. We obtain 
(assuming $N>2$ and $\tilde{N}>2$)
\begin{align}
G_{m=3,l=2} (x,\xi,\eta)&= (a-\xi) (a-\eta) G_{FS}(x)\nonumber \\ 
&-\sum_{a=1}^N A_a\ (a-\xi_a)(\eta-\xi_a)(\xi-\xi_a)\log (\xi-\xi_a) \\
&-\sum_{a=1}^{\tilde{N}} B_a\ (a-\xi_a)(\xi-\eta_a)(\eta-\eta_a)
\log (\eta-\eta_a) 
 \nonumber\ ,
\end{align}
where $A_a$ and $B_a$ are as defined in \eqref{ABdefs}. Again, 
note the appearance of $p(t)$ in the coefficient of the 
logarithms and the coefficient of the Fubini-Study metric is 
simply $p(t)$ without the constant root -- this is called 
$p_{nc}(t)$ in \cite{ACG1}.

\subsection{The $m=3$, $l=1$ symplectic potentials}

\subsubsection*{For momentum polynomial $p_1(t)=(t-a)^2(t-\chi)$}

Let $p_{nc}(t)=(t-\chi)$ be the part of the momentum polynomial 
involving the non-constant root $\chi$. Further, let us assume 
that $h$ is a polynomial of degree $N$ with distinct roots 
$\chi_1,\ldots,\chi_N$. Then we can write 
$h(\chi)=-h_0\prod_{r=1}^N(\chi-\chi_r)$. Then, the symplectic 
potential takes the form
\begin{equation}
\Big[G_{m=3,l=1}\Big]_{p_1(t)} = 
-\sum_{r=1}^N  C_r\ p_1(\chi_r)\ \log (\chi -\chi_r) +
p_{nc}(a)\ G_a\ ,
\end{equation}
where $C_r\equiv -h_0^{-1}\prod_{s\neq r}(\chi_r-\chi_s)^{-1}$ and 
$G_a$ is the symplectic potential for the small metric $ds^2_a$.

\subsubsection*{For momentum polynomial $p_2(t)=(t-a)(t-b)(t-\chi)$}

The symplectic potential takes the form ($p_{nc}(t)=(t-\chi)$)
\begin{equation}
\Big[G_{m=3,l=1}\Big]_{p_2(t)} = 
-\sum_{r=1}^N  C_r\ p_2(\chi_r)\ \log (\chi -\chi_r) +
p_{nc}(a)\ G_a + p_{nc}(b) G_b \ ,
\end{equation}
where we have again assumed that $h(\chi)$ is a polynomial of 
degree $N$. Further $G_a$ (resp. $G_b$) is the symplectic 
potential for the small metric $ds^2_a$ (resp. $ds^2_b$).

\subsection{Global Analysis of the ACG metrics}

While most of our analysis will hold in generality, we will 
restrict all our considerations to the situation when the 
functions $f$ and $g$ are cubic functions. Let
\begin{equation}
f(\xi)=-f_0 (\xi-\xi_1)(\xi-\xi_2)(\xi-\xi_3) \quad,\quad
g(\eta) =-g_0 (\eta-\eta_1)(\eta-\eta_2)(\eta -\eta_3) \ ,
\end{equation}
with the ordering $\xi_1<\xi_2<\xi_3$ and $\eta_1<\eta_2<\eta_3$ 
when the roots are all real. If $f$ has complex roots, we choose 
them to be $\xi_2$ and $\xi_3$ and similarly for the function 
$g$. We also will assume $f_0$ and $g_0$ are real and positive. 
We will now consider the various ACG metrics and require that the 
metric be positive-definite.
\begin{itemize}
\item[] \hspace{-27pt}\textbf{$m=2$ orthotoric metrics:}  
The following conditions are needed
$$
\xi > \eta \quad, \quad \xi < \xi_1 \textrm{ or }  \xi_2 
< \xi < \xi_3  \quad,\quad
\eta_1 < \eta < \eta_2 \textrm{ or } \eta > \eta_3\ ,
$$
and a similar set of conditions if we require $\xi < \eta$. If we 
require that the four-dimensional space be compact, clearly, we 
obtain the condition that $\xi_2 > \eta_2$ thus satisfying 
$\xi>\eta$ everywhere in the interior. The metric is singular on 
the boundary of a rectangle in the $\xi-\eta$ plane. These 
metrics lead to $\BC\BP^2$ when the functions $f$ and $g$ are 
identical\cite{Acharya:2006as}.
\item[] \hspace{-27pt}\textbf{$m=3$, $l=1$ ACG metrics:} Let us 
assume that the non-compact coordinate is $\chi$. Then, 
positivity of the metric is assured when $\chi<\chi_1<a$, where 
$\chi_1$ is the smallest root of $h(\chi)$. We choose the 
four-dimensional metric to be the one given by the $m=2$ 
orthotoric metric. Again, the singularities are given by those of 
the $m=2$ orthotoric metric that we just considered and at 
$\chi=\chi_1$. These metrics will be shown to lead to complex 
cones over $L^{pqr}$ spaces when $f\neq g$ for specific choices 
of the polynomials.
\item[] \hspace{-27pt}\textbf{$m=3$ $l=2$ ACG metrics:} The 
relevant conditions are
$$
\xi< a \ ,\quad \eta <a \ ,\quad \xi > \eta \ , \quad 
 \xi_1 < \xi < \xi_2 \textrm{ or } \xi > \xi_3 \ ,\quad
\eta < \eta_1 \textrm{ or }  \eta_2 < \eta < \eta_3\ ,
$$
and of course, $-1<x<1$. In this example, we will be interested 
in the situation when we have non-compact domain in the 
$\xi-\eta$ plane given by $ \eta < \eta_1 < \xi_1<\xi < \xi_2< 
a$. The singularities of the metric occur at $x=\pm 1$ (these are the 
singularities of the FS metric), 
$\eta=\eta_1$, $\xi=\xi_1,\xi_2$. This choice leads to partially resolved 
cones over $Y^{pq}$ as we will show in the sequel.
\end{itemize} 

\subsection{Recovering the canonical potential}

Consider the simple situation of the $m=2$ orthotoric symplectic 
potential corresponding to $f=A_1/(\xi-\xi_1)$. Ignoring the 
non-logarithmic pieces, the symplectic potential given in 
\eqref{simplepart} can be re-written as
\begin{equation}
\frac12 \ell(P) \log \ell(P) +A_1(\xi-\xi_1)(\eta -\xi_1) \log (\eta -\xi_1)
\end{equation} 
where $\ell(P)=-2A_1(\xi_1 P_1 -P_2 -\xi_1^2)$. The singularity 
associated with $\ell(P)=0$ is split into two separate 
singularities in the $\xi-\eta$ space i.e., $\xi=\xi_1$ and 
$\eta=\xi_1$. The ACG metric \textit{subtracts} out one of the 
two singularities and hence has the form $G_{\textrm{can}}+h(P)$. 
This simple example shows how one can re-write all the symplectic 
potentials that we have obtained into the form
$$
G_{\textrm{can}} + h(P) \ ,
$$
where $G_{\textrm{can}}$ contains only the $\ell \log \ell $ 
pieces associated with the singularities that we obtained from 
our global analysis. All the remaining pieces are grouped 
together into the Abreu function $h(P)$. We can then use the 
canonical potential to figure out the moment polytope.

\section{Unresolved Ricci-flat metrics:~examples}

We consider the $m=3$, $l=1$ ACG metrics with momentum polynomial 
$p_1(t)$. We define $P=(a-\chi)$ and choose the cubic polynomial 
to be $h(\chi)=2(a-\chi)^3$. Then the metric in Eqn. 
\eqref{m3l1a} is the metric for the complex cone over a 
four-dimensional K\"ahler manifold. The symplectic potential then 
takes the form \begin{equation} G_{m=3,l=1} = \frac12 P \log P + 
P\ G_a(\tilde{P}_1,\tilde{P}_2)\ , \end{equation} where 
$\tilde{P}_i\equiv P_i/P$ and $G_a$ is the symplectic potential 
for a four-dimensional manifold which we take to be $m=2$ 
orthotoric manifold. Thus, we have $G_a=G_{\textrm{OT2}}$ as 
defined in Eqn. \eqref{sympotOT2}. The Ricci-flatness of the 
above metric requires $G_a(\tilde{P}_1,\tilde{P}_2)$ to be 
K\"ahler-Einstein with scalar curvature equal to $4$ among other 
things. This is achieved if we choose $f'(x)=g'(x)$ and 
$f_0=g_0=\tfrac23$.

Focusing on Einstein spaces amongst the ACG metrics in four 
dimensions, we thus need to consider cubic function $f$ and $g$ such 
that
\begin{equation}\label{fglpqr}
f(x)=- \tfrac23 x^3 + f_1 x^2 + f_2 x + f_3\ ,\quad g(x)=f(x)+\mu \ ,
\end{equation} 
with $\mu\neq0$ and $(f_1,f_2,f_3)$ are constants.
Let $\xi_1< \xi_2 < \xi_3$ be the distinct 
real roots of $f$  and $\eta_1 < \eta_2 < \eta_3$ be  distinct real 
roots of $g$. 
As discussed earlier, we choose the values of $\xi$ and $\eta$ 
are such that $\eta_1 < \eta < \eta_2 < \xi_2 < \xi < \xi_3$
This implies that the singularities occur on the boundary of a rectangle
in the $\xi-\eta$ plane. In the $(\tilde{P}_1=\xi+\eta,\tilde{P}_2=\xi\eta)$ 
plane, the rectangle is  given by the conditions $\ell_a=0$ where
\begin{eqnarray}
\ell_1(\tilde{P}_1,\tilde{P}_2) &=
\frac{-2}{f_0(\eta_1-\eta_2)(\eta_1-\eta_3)}(\eta_1 \tilde{P}_1 -\tilde{P}_2
-\eta_1^2)\ , \nonumber \\
\ell_2(\tilde{P}_1,\tilde{P}_2) &=
\frac{-2}{f_0(\eta_2-\eta_1)(\eta_2-\eta_3)}(\eta_2 \tilde{P}_1 -\tilde{P}_2
-\eta_2^2)\ , \nonumber \\
\ell_3(\tilde{P}_1,\tilde{P}_2) &=
\frac{-2}{f_0(\xi_2-\xi_1)(\xi_2-\xi_3)}(\xi_2 \tilde{P}_1 -\tilde{P}_2 -\xi_2^2)\ ,\\
\ell_4(\tilde{P}_1,\tilde{P}_2) &=
\frac{-2}{f_0(\xi_3-\xi_1)(\xi_3-\xi_1)}(\xi_3 \tilde{P}_1
-\tilde{P}_2 -\xi_3^2)\  \nonumber .
\end{eqnarray}
The four functions are linearly-dependent. We assume that the dependence is 
given by
four positive integers $(p,q,r,s)$ such that (assuming $q>p$ and $s>r$)
$$
p\ \ell_1(\tilde{P}_1,\tilde{P}_2) + q\ \ell_2(\tilde{P}_1,\tilde{P}_2)
- s\ \ell_3(\tilde{P}_1,\tilde{P}_2) - r\ \ell_4(\tilde{P}_1,\tilde{P}_2)=0\ ,
$$
The condition $p+q=r+s$ may be assumed at this point but it can 
be obtained as a consistency condition. For simplicity, we assume 
that this is true. These spaces turn out to be real cones over 
five-dimensional spaces called $L^{pqr}$\cite{Cvetic:2005ft}. The 
explicit map relating this $m=3,l=1$ ACG metric to metrics given 
in \cite{Cvetic:2005ft} has been obtained by Martelli and 
Sparks(in \cite{Martelli:2005wy}) and we shall not present them 
here. We instead pursue our analysis to completion. The four 
integers should determine the functions $f$ and $g$. Using the 
above condition we obtain the three equations after setting 
$\eta_1=0$ by a simultaneous translation in $\xi$ and $\eta$ and 
choosing $\eta_2=1$ for simplicity as the $\eta_2$-dependence can 
be easily recovered:
\begin{eqnarray}
\label{rels}
p B_1 +q B_2 - s A_2 - r A_3=0\ , \nonumber \\
q B_2 -s A_2 \xi_2 -r A_3 \xi_3=0\ , \\
q B_2 -s A_2 \xi_2^2 -r A_3 \xi_3^2=0\ , \nonumber
\end{eqnarray}
where $B_1=1/\eta_3$, $B_2=1/(1-\eta_3)$, $A_2=1/(\xi_2-\xi_1)(\xi_2-\xi_3)$
and $A_3=1/(\xi_3-\xi_1)(\xi_3-\xi_1)$. Note that it seems that we have
four variables to determine, $\eta_3$ and the three roots of $g$. However,
the two functions $f$ and $g$ are such that their roots satisfy
$$
\xi_1+\xi_2+\xi_3 =\eta_1+\eta_2+\eta_3\ ,\textrm{and}\ 
\xi_1\xi_2+\xi_3\xi_1+\xi_2\xi_3 =\eta_1\eta_2+\eta_3\eta_1+\eta_2\eta_3\ .
$$
This enables us to solve for, say, $\eta_3$  and $\xi_1$ in terms of 
$\xi_2$ and $\xi_3$ to obtain:
\begin{equation}
\eta_3=1-\frac{\xi_2\xi_3}{\xi_2+\xi_3-1}\quad \textrm{and} \quad
\xi_1=(\xi_2+\xi_3)-\frac{\xi_2\xi_3}{\xi_2+\xi_3-1}\ .
\end{equation}
Thus, Eqn. (\ref{rels}) now becomes three  equations for two variables, 
$\xi_2$ and $\xi_3$ given four integers $(p,q,r,s)$ such that
$p+q-r-s=0$.

One can also view Eqn. (\ref{rels}) as an equation for three rational
numbers $\tfrac1p(q,r,s)$ given $\xi_2$ and $\xi_3$. One can show
that the solution is such that $p+q-r-s=0$ and one has
\begin{eqnarray}\label{pqtouv}
\frac{q-p}{q+p}&=&
\frac{( u + v ) 
      ( u^2 + v^2-2 ) }{uv(  u^2 + uv + v^2 -2) -1} > 0\ ,
\nonumber \\
\frac{s-r}{p+q} &=&
\frac{( v-u ) ( u + v ) 
    ( 1 + uv ) }{uv(  u^2 + uv + v^2-2 )-1 } >0\ ,
\end{eqnarray}
where we have defined
\begin{equation*}
\xi_2 =\frac{1+u}2\quad,\quad
\xi_3 =\frac{1+v}2\quad \textrm{with } v>u>1\ .
\end{equation*}
The above range of $(u,v)$ is consistent with the condition that 
$q>p$ and $s>r$ that we assumed at the beginning. It is easy to 
see that when $u$ and $v$ are rational, one is guaranteed to 
obtain integers for $(p,q,r,s)$. This solution is similar to the 
one considered in \cite{Cvetic:2005vk}.

Consider the example when $(p,q,r,s)=(1,4,2,3)$. We solve for 
$(u,v)$ numerically as the explicit answers are unilluminating. 
We obtain that $(u,v)=(1.8933,2.3258)$ -- this is the only 
solution that satisfies $v>u>1$. This implies that 
$\eta_3=1.96867$ and 
$(\xi_1,\xi_2,\xi_3)=(-0.14065,1.44665,1.66229)$. Again this is 
consistent with the ordering of the roots that we assumed.

It is of interest to ask what happens when $r=s$ in 
\eqref{pqtouv}. It is not hard to see that this is achieved when 
$u=v$. In other words, one has $\xi_2=\xi_3$ and one side of the 
rectangle shrinks to zero size. The root $\xi$ now becomes a 
constant root. The singularity may be `resolved' by associating a 
$\BC\BP^1$ with the constant root. This provides an intuitive 
understanding of our next attempt to find metrics for $Y^{pq}$ 
from the $(m=3,l=2)$ ACG metrics.

\section{Partially resolved Ricci-flat metrics:~examples}

\subsection{Cones over $Y^{pq}$ spaces}

The toric data for general $Y^{p,q}$ (with $p>q$ and gcd$(p,q)=1$) is
given by the four vectors\cite{Martelli:2006yb}
\begin{displaymath}
\mathcal{V}=
\begin{pmatrix}
1 & 1 & 1 & 1  \\ 0 & p-q-1 & p & 1 \\ 0 & p - q & p & 0   \  
\end{pmatrix}\ .
\end{displaymath}
One can verify that the most general internal point is of the 
form $(1,k,k)$ with $k \in \{1, \ldots, (p-1)\}$. Internal points 
correspond to blowing up four-cycles and we intend to add one 
internal point and obtain the Ricci-flat metric on the resulting 
space. Now, with one internal point added, the toric data is
\begin{displaymath} 
\mathcal{V}_{+1}=
\begin{pmatrix}
1 & 1 & 1 & 1 &1 \\ 0 & p - q -1  & p & 1 & k\\ 0 & p - q & p & 0  & k \  
\end{pmatrix}\ .
\end{displaymath}

The general D-terms for $Y^{p,q}$ spaces with one internal point added can 
be computed by considering the null space to $\mathcal{V}_{+1}$ and
turns out to be  
\begin{eqnarray}\label{YpqDterm}
	(p-q)\ell_1 + (p+q) \ell_3 - p(\ell_2 + \ell_4) &=& r_1\ , \nonumber \\
	\left( -k + p \right) \,{\ell_1} + k {\ell_3} - p\,{\ell_5} &=& r_2\ ,
\label{dterm}
\end{eqnarray}
where we have also turned on the blow-up (F-I) parameters which we call
$r_1$ and $r_2$. We thus have the 
five $\ell_a$ being subject to these two conditions. This 
effectively leaves us with three independent fields. We choose 
these independent fields to be $(P_1, P_2, P_3)$.

The metrics for $Y^{pq}$ spaces were first obtained in 
\cite{Gauntlett:2004yd,Gauntlett:2004hh}. Real cones over these 
spaces have a conical singularity at the tip of the cone. 
Resolved metrics for these spaces have been not been found except 
for the conifold (and its $\BZ_2$ orbifold). The conifold is 
obtained as a real cone over $Y^{1,0}=T^{1,1}$. An intriguing 
result was obtained in \cite{Oota:2006pm} where they obtained a 
resolved metric for the cone over $Y^{2,1}$. What was different 
about this result was the fact that the blow-up parameters were 
set to fixed values. We realised that the metric looked like the 
$m=3$, $l=2$ ACG metric and verified that it was indeed true.  
This was our inspiration to look more closely at this class of 
ACG metrics and see if we could achieve similar results for 
general $Y^{p,q}$. Further,  the defining D-term for cones over 
$Y^{p,q}$ spaces clearly has a $\BC\BP^1$ corresponding to 
$\ell_2+\ell_4=$constant in Eqn. \eqref{YpqDterm}. We now 
systematically fit the $m=3$, $l=2$ ACG metrics to the two 
D-terms that appear in Eqn. \eqref{YpqDterm}. Higher dimensional 
generalisations of the result of \cite{Oota:2006pm} have appeared 
in \cite{Lu:2006cw}. Our result provides examples in six 
dimensions that appear to be new.

\subsection{Fitting to the $m=3$, $l=2$ ACG metrics}

We now attempt to fit these metrics in to the $m=3$, $l=2$ class 
of ACG metrics. We first set $ds^2_a$ to the Fubini-Study metric 
normalised such that the scalar curvature equals 2. As discussed 
earlier, Ricci-flatness requires $f$ and $g$ to be cubic 
functions such that
\begin{equation}
f'(x)=g'(x)=(x-a)(b_1 x + b_2)\ ,
\end{equation}
with $b_1 a + b_2=-R(ds_a^2)=-2$.  A simultaneous shift in $\xi$ 
and $\eta$ can be done to eliminate the term linear in $x$ that 
appears in the functions $f$ and $g$. This is achieved, for 
instance, by setting $b_2=0$. We also set $a=1$ to match results 
in the literature. This fixes $b_1=-2$. Thus we obtain
\begin{equation}
f'(x)= -2 x(x-1) \implies f(x)=-\tfrac23 x^3 + x^2 + \textrm{constant}\ ,
\end{equation}                      
and $g(x)-f(x)$ is a constant.

We identify $(P_1,\hat{P}_2,P_3)$ with an $SL(3,\BZ)$ transform 
of the coordinates given in \eqref{Ypqcoords}. The $SL(3,\BZ)$ 
transform is such that $P_2 = (\hat{P}_2+P_1-1)$ leaving the 
other two coordinates unchanged. To carry out the fit to the 
$Y^{pq}$ D-terms, we identify the five singularities of the $m=3$ 
$l=2$ ACG metric with boundary of the $Y^{pq}$ polytope. The 
singularities of the $\BC\BP^1$ are naturally identified with 
$\ell_2$ and $\ell_4$. We find that the $\xi=\xi_1$ and 
$\xi=\xi_2$ singularities get identified with the $\ell_1$ and 
$\ell_3$ singularities. If the fit has to work, the last 
singularity $\eta=\eta_1$ \textit{must} be identified with 
$\ell_5=0$ singularity. With these inputs, we obtain
\begin{eqnarray}
	\ell_1 &=& -(1-\xi_1)A_1 \bigg[P_1 (\xi_1 -1) -\hat{P}_2 + 
1-\xi_1^2 \bigg]\ ,\nonumber\\
	\ell_2 &=& (\hat{P}_2 + P_3) \ ,\nonumber \\
	\ell_3 &=& -(1-\xi_2)A_2 \bigg[ P_1 (\xi_2 -1) -\hat{P}_2 + 
1-\xi_2^2 \bigg] \ ,\\
	\ell_4 &=& (\hat{P}_2 - P_3)\ , \nonumber\\
	\ell_5 &=& -(1-\eta_1)B_1 \bigg[ P_1 (\eta_1 -1) -\hat{P}_2 + 
1-\eta_1^2 \bigg]\ ,\nonumber
\label{lis}
\end{eqnarray}
where $\xi_i$ and $\eta_i$ are respectively the roots of cubic 
equations $f(\xi)=0$ and $g(\eta)=0$.  The roots are taken to 
have the following ordering: $\eta<\eta_1<\xi_1<\xi<\xi_2$. The 
constants $A_1$, $A_2$ and $B_1$ are given by
\begin{equation}
	A_1 = \frac{-3 }{(\xi_1 - \xi_2)(\xi_1-\xi_3) } \ ,\quad
	A_2 = \frac{-3 }{(\xi_2 - \xi_1)(\xi_2-\xi_3) } \ ,\quad
	B_1 = \frac{-3 }{(\eta_1 - \eta_2)(\eta_1-\eta_3) } \  .
\end{equation}

We now need to impose the conditions that the $\ell_a$ as given 
above from the $m=3$, $l=2$ ACG metric satisfies the D-term 
conditions given in Eqn. \eqref{YpqDterm}. In the first D-term, 
one sees that the $P_3$ drops out and thus leads to two equations 
corresponding to the vanishing of the coefficients of $P_1$ and 
$\hat{P}_2$. Further, this does not involve the roots of $g$ since 
they appear only in $\ell_5$. Here $f$ is such that its roots satisfy
$$
\xi_1+ \xi_2 + \xi_3  =\frac32 \quad,\quad
\xi_1 \xi_2 +\xi_2 \xi_3 + \xi_3 \xi_1 =0\ .
$$
Thus, the first D-term is an over-determined system -- we have 
two equations and one unknown -- the undetermined constant in 
$f$. It turns out there is indeed a solution.
\begin{eqnarray}
\xi_1 = \frac{2\,p - 3\,q - {\sqrt{\Delta}}}{4p} \ ,\quad
	\xi_2 = \frac{2\,p + 3\,q - {\sqrt{\Delta}}}{4p} \ ,\quad
\xi_3 =\frac{p + {\sqrt{\Delta}}}{2p}\ ,
\end{eqnarray}
where $\Delta\equiv (4p^2 - 3q^2)$. It is easy to verify that the 
inequality $\xi_1<\xi_2<1$ is satisfied when $p>q$. Note that 
$\xi_i$ are \textit{independent} of $k$, i.e., the interior point 
that is blown-up. This is obvious since the second D-term was not 
used in determining the roots of $f$. The FI parameter $r_1$ is 
non-vanishing and is given by
\begin{equation}
r_1=-\frac12\left(\frac{2p^2-3q^2}p + \sqrt{\Delta} \right)\ .
\end{equation}
It turns out that $r_1$ is always negative. This is consistent 
with our identification of the $\BC\BP^1$ arising with from the 
$\ell_2+\ell_4$.

We now impose the second D-term equation involving $\ell_5$ and 
use it to determine the roots of $g$. Again, we know that the 
three roots of $f$ must satisfy
\begin{eqnarray}
	\eta_1 +\eta_2 + \eta_3=\tfrac32 \quad,\quad
	\eta_1 \eta_2 +\eta_2 \eta_3 + \eta_3 \eta_1 =0\ .
\end{eqnarray}
We can use these two equations to solve for $\eta_2$ and $\eta_3$ 
in terms of $\eta_1$. Imposing the D-term leads to the solution
\begin{eqnarray}
	\eta_1 = \frac{p( 2p - 3q) (p + q)  - 2k( 2p^2 - 3q^2 )  
- \sqrt{\Delta}( -2kp + p(p + q) 
       ) }{4( 3k^2q + p^2(p + q)  - kp( 2p + 3q)) }
\end{eqnarray}
The second FI parameter is given by 
\begin{equation}
r_2=	-(p -k ) (1-\xi_1)A_1 ( 1-\xi_1^2) - k (1-\xi_2)
A_2(1-\xi_2^2) +p (1-\eta_1)B_1 (1-\eta_1^2)  
\end{equation}
We do not list the explicit expressions for $\eta_2$, $\eta_3$ as we 
don't really need them. Instead, we just note the value of 
their sum and product since they appear directly in $B_1$ which appears
in $\ell_5$.
\begin{eqnarray}
	\eta_2 + \eta_3 = \tfrac{3}{2} - \eta_1\ ,\quad
	\eta_2 \eta_3 = -\eta_1 (\tfrac{3}{2} -\eta_1)\ .
\end{eqnarray}
The constants that appear in $f$ and $g$ can be obtained directly 
from the roots and we do not give expressions for them.

An important point to note here is that we have not verified that 
$\eta_1<\xi_1$ as that is required by the positivity of the ACG 
metric. While our expressions seem to be valid for any  $k\in 
(1,2,\ldots, p-1)$, it turns out that in all the examples that we 
have considered, the inequality is violated when $k=p-1$ and 
$p>2$. Experimentally, we find that for all values of $k$ that 
are greater than $p/2$ and thereabouts, the inequality is 
violated and we do \textit{not} obtain a resolved metric for 
those values of $k$. For instance, for $Y^{3,1}$, we obtain a 
resolved metric for $k=1$ but the one for $k=2$ violates the 
inequality and we do not have a positive definite metric.

The Abreu function may be extracted using the formula 
\begin{eqnarray}
	G_{m=3,l=2} &=& \frac{1}{2}\sum_{a=1}^{5} \ell_a 
\log[\ell_a] + h(P_i) \ .
\end{eqnarray}
We do not write an explicit formula for the Abreu function. We 
now work out details for some specific values of $(p,q,k)$. The 
polynomials are taken to be\footnote{Our metrics differ from the 
ones usually written for 
$Y^{pq}$\cite{Gauntlett:2004yd,Gauntlett:2004hh,Oota:2006pm} by a 
factor of 3 due to our choice of normalisation for the scalar curvature 
of $\BC\BP^1$. Our $f$ will have to be multiplied by $-3$ to 
match with the corresponding cubic function in those papers.}
\begin{eqnarray}
	f(\xi) = -\tfrac23 \xi^3 + \xi^2 + a \ ,\quad
	g(\eta) = -\tfrac23 \eta^3 + \eta^2 + b\ ,
\end{eqnarray}
where we define the constants to be $a$ and $b$ (this is not to 
be confused with our earlier use of the same to indicate constant 
roots in the momentum polynomial).

\subsubsection*{$\mathbf{Y^{2,1}}$}

There is only one point in the interior of the polytope 
corresponding to setting $k=1$. The FI parameters are given by 
\begin{eqnarray}
	r_1= -\tfrac{1}{4} (5 +2 \sqrt{13})\ ,\quad
	r_2= \tfrac{3}{4} (4 + \sqrt{13}) \ .
\end{eqnarray}
\begin{figure}[h]
\begin{center}
	\includegraphics[scale=0.3]{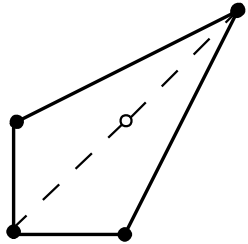}
	\caption{Toric diagram for $Y^{2,1}$}
	\label{y21}
\end{center}
\end{figure}
The form of the resolved $Y^{2,1}$ metric in the 
\cite{Oota:2006pm} can be obtained by carrying out an $SL(3,\BZ)$ 
transformation such that the new $P_i$ are given by
\begin{eqnarray}
	P_1 = \xi + \eta \ , \quad
	P_2 = (1-\xi)(1-\eta)\ , \quad
	P_3 = (1-\xi)(1-\eta)x\ ,
\label{transformation}
\end{eqnarray}
and setting $x=\cos\theta$. The roots of the polynomials turn out to be
\begin{eqnarray}
	\xi_1 = \frac{1}{8}(1-\sqrt{13}) \ , \ 
	\xi_2 = \frac{1}{8}(7- \sqrt{13})  \ , \ 
	\xi_3 = \frac{1}{4}(2 + \sqrt{13}) \ , \nonumber \\
	\eta_1 = -\frac{1}{2}(2 + \sqrt{13}) \ , \ 
\eta_2 = \bar{\eta}_3 = \frac{5 + {\sqrt{13}} - i 
\,{\sqrt{54 + 18\,\sqrt{13}}}}{4}\ .
\end{eqnarray}
The constants $a$ and $b$ appearing in the functions $f$ and $g$ are 
\begin{eqnarray}
	a = - \frac{1}{96}(16 - \sqrt{13}) \ ,\quad
	b= -\frac{1}{12}(137 + 37\sqrt{13})\ .
\end{eqnarray}

\subsubsection*{$\mathbf{Y^{3,1}}$ with $\mathbf{k=1}$}

\begin{figure}[h]
\begin{center}
	\includegraphics[scale=0.3]{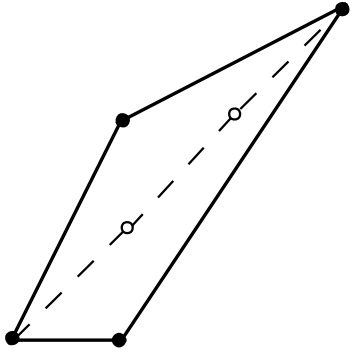}
	\caption{Toric diagram for $Y^{3,1}$}
	\label{y31}
\end{center}
\end{figure}

For $Y^{3,1}$, we have two internal points. We add 
the point corresponding to the vector $(1,1,1)$.  The other point 
$(1,2,2)$ does \textit{not} give a positive definite metric and 
hence is not considered. The roots when $k=1$ turn out to be
\begin{eqnarray}
	\xi_1 = \frac{3 - \sqrt{33}}{12}\  , \ 
	\xi_2 = \frac{9- \sqrt{33}}{12}\   , \
	\xi_3 = \frac{ 3 + \sqrt{33}}{6}\  , \
	\eta_1 = \frac{1 - \sqrt{33}}{8}\ .
\end{eqnarray}
We note that the ordering $\eta_1 < \xi_1 < \xi_2 < \xi_3$ is respected.
The constants in the two polynomials are
\begin{eqnarray}
a =  -\frac{9 -\sqrt{33}}{54}  \ , \quad b = -\frac{77 + 
3\sqrt{33}}{192}\ .
\end{eqnarray}
The FI parameter $r_2=\tfrac18(7+\sqrt{33})$.

\subsubsection*{$\mathbf{Y^{3,2}}$ with $\mathbf{k=1}$}

\begin{figure}[h]
\begin{center}
	\includegraphics[scale=0.3]{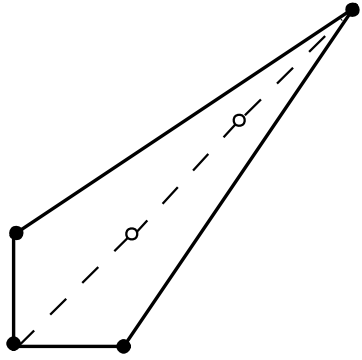}
	\caption{Toric diagram for $Y^{3,2}$}
	\label{y32}
\end{center}
\end{figure}

This example is similar to $Y^{3,1}$. The $k=2$ solution is not valid 
but the $k=1$ is and hence we present the results for that metric. 
The roots turn out to be 
\begin{eqnarray}
	\xi_1 = -\frac1{\sqrt{6}}  \ ,\ 
	\xi_2 = \frac{6- \sqrt{6}}{6}  \ ,\ 
	\xi_3 = \frac{ 3 + 2 \sqrt{6}}{6} \ ,\
	\eta_1 = - \frac{2 + 3\sqrt{6}}{10} \ .
\end{eqnarray}
We note that the ordering of the roots is as expected.
The constants that appear in the polynomials are
\begin{eqnarray}
	a=- \frac{ 9 + \sqrt{6}}{54}\ , \quad
	b= -\frac{601 + 189{\sqrt{6}}}{750}\ .
\end{eqnarray}
The FI parameter $r_2=\tfrac25(4+\sqrt{6})$.

\subsubsection*{$\mathbf{Y^{5,3}}$ with $\mathbf{k=1,2}$}

\begin{figure}[h]
\begin{center}
	\includegraphics[height=2.5in]{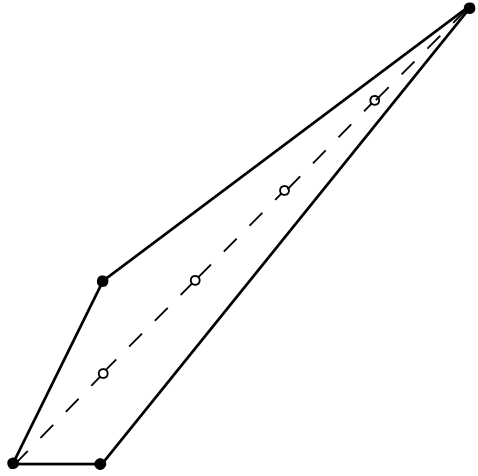}
	\caption{Toric diagram for $Y^{5,3}$}
	\label{y53}
\end{center}
\end{figure}

This is the first example where we obtain inequivalent 
resolutions corresponding to adding internal points for $k=1$ and 
$k=2$. Since the roots of $f$ are independent of $k$, we will 
quote them once and write out the root $\eta_1$ separately. We 
obtain
\begin{eqnarray}
	\xi_1 =\frac{1- \sqrt{73}}{20}  \ ,\ 
	\xi_2 = \frac{19- \sqrt{73}}{20}  \ ,\ 
	\xi_3 = \frac{ 5 +  \sqrt{73}}{10} \ ,\ \nonumber \\
	\eta_1(k=1) = - \frac{1 + 5\sqrt{73}}{76} \ , \
	\eta_1(k=2) = - \frac{13 + 5\sqrt{73}}{46}\ .
\end{eqnarray}
We note that the ordering of the roots is as expected.
The constants that appear in the polynomials are
\begin{eqnarray}
	&a=- \frac{ 125 + \sqrt{73}}{750}\ ,&\nonumber\\ 
	&b(k=1)= -\frac{26705+ 1285{\sqrt{73}}}{82308}\ ,\ \ 
	b(k=2)= -\frac{105479 + 10315{\sqrt{73}}}{73002}\  .&
\end{eqnarray}
The FI parameter $r_2(k=1)=\tfrac3{190}(77+5\sqrt{73})$ and 
$r_2(k=2)=\tfrac{27}{460}(59+5\sqrt{73})$.

One can ask what happens to our formulae when $q=0$. The roots 
$\xi_1$ and $\xi_2$ coincide. This implies that the $m=3$, $l=2$ 
ACG metric is singular. $\xi$ becomes a constant root. This is 
similar to what happened in the $L^{pqr}$ metric earlier. Again, 
we need to add a $\BC\BP^1$ to resolve this singularity. So it 
naturally leads us to the conifold and its orbifolds. We thus 
move on to the $m=3$, $l=1$ ACG metrics with momentum polynomial 
$p_2(t)$.

\subsection{The $m=3$, $l=1$ ACG metric and the resolved conifold}

The metric for the resolved conifold as well its $\BZ_2$ orbifold 
has been obtained by in \cite{PandoZayas:2000sq} and 
\cite{PandoZayas:2001iw}. Following these papers, both the 
metrics can be written as\footnote{Below $ds^2_{\BC\BP^1}$ is the 
metric $(d\theta^2+\sin^2\theta d\varphi^2)$ and 
$\omega=\cos\theta d\theta\wedge d\varphi$ is the K\"ahler form 
for $\BC\BP^1$. The indices $a$ and $b$ distinguish the two 
$\BC\BP^1$'s that appear.}
\begin{equation}
ds_6^2 = \kappa^{-1}(\rho)d\rho^2 + \tfrac{\rho^2}9 \kappa(\rho) 
(d\psi-A_a-A_b)^2 +
\tfrac{\rho^2}6  ds^2_{\BC\BP_a^1}+\big(\tfrac{\rho^2}6 +\hat{a}^2\big) 
ds^2_{\BC\BP_b^1}
\end{equation}
where 
$$
\kappa(\rho)=(1+\tfrac{9\hat{a}^2}{\rho^2}
-\tfrac{\hat{b}^6}{\rho^6})/(1+\tfrac{6\hat{a}^2}{\rho^2})\ , \ dA_a=\omega_a\ 
\textrm{and}\ dA_b=\omega_b\ .
$$
The metric of the resolved conifold is obtained after setting 
$\hat{b}=0$ and choosing the periodicity the angle $\psi$ to be 
$4\pi$. The metric of the resolution of the $\BZ_2$ orbifold of 
the conifold is obtained by simply choosing the periodicity of 
$\psi$ to be $2\pi$. The periodicity of the angles are determined 
by requiring the metrics to be non-singular at $\rho=0$. The 
parameter $\hat{a}$ is the size of the blown-up $\BC\BP^1$. We 
will now show how these two metrics are indeed $m=3$, $l=1$ ACG 
metrics with momentum polynomial $p_2(t)$.

Hence consider the $m=3$, $l=1$ ACG metrics and choose $ds^2_a$ and 
$ds^2_b$ to be the Fubini-Study metric on $\BC\BP^1$. Both are 
taken to have scalar curvature equal to $2$. Ricci-flatness of the 
metric implies that the function $h(\chi)$ must be of the form
\begin{equation}
h'(x)=-2 (x-a)(x-b)\ ,
\end{equation}
With no loss of generality, assume $a=0$ and $b>0$. Then, we obtain
\begin{equation}
h(x) =-\frac23 x^3 + b x^2 - c\ ,\quad c=\textrm{a positive constant}\ .
\end{equation}
The resolved conifold is obtained when the constant $c=0$. 
Positivity of the metric requires $\chi<0$. The metric that we 
obtain here becomes the metric on the resolved conifold given in 
\cite{PandoZayas:2000sq} after the change of variable 
$\chi=-\rho^2/6$, identifying $\theta_1=(d\psi-A_a-A_b)$ and 
finally setting $b=\hat{a}^2$.

The metric for the $\BZ_2$ orbifold is obtained if we choose the 
constant $c=\hat{b}^6/324$ and choose the periodicity of $\psi$ to be 
$2\pi$ as follows from the analysis in \cite{PandoZayas:2001iw}.

\subsection{The resolution of $\BC^3/\BZ_3$ as a $l=3$, $m=1$ ACG
metric}

The metric for the resolution of $\BC^3/\BZ_3$ when written as
the resolution of a cone is(after a rescaling)\cite{Aswinthesis}
\begin{equation}
ds^2=3\bigg[ 1 - \frac{b^6}{r^6} \bigg] ^{-1} dr^2 + \frac{r^2}3
\bigg[1 - \frac{b^6}{r^6} \bigg] \bigg(d\gamma +  A\bigg)^2 +
\frac{r^2}2 ds_{\mathbb{P}^2}^2 \ .
\end{equation}
where $ds_{\mathbb{P}^2}^2$ is the Fubini-Study metric on $\mathbb{CP}^2$
with scalar curvature equal to $8$ and the K\"ahler form is
$\omega_{\mathbb{P}^2}\equiv -dA$.

We wish to show that this is an example of the $m=3$ $l=1$ with 
momentum polynomial $p_1(t)$. We take the small K\"ahler metric 
to be the Fubini-Study metric on $\BC\BP^2$ with scalar curvature 
equal to $8$. The Ricci-flatness condition requires $h(\chi)$ 
such that $h'(x)=b_1(x-a)^2$ with $b_1 =-4$. Setting the constant 
$a$ to zero with no loss of generality, we obtain
$$
h(\chi)=-\frac43 \chi^3 -\frac{b^6}6\ ,
$$
where we have chosen the constant suitably.
One further has the condition that $d\theta_1=-\omega_{\BP^2}$.
Identifying $\chi=-r^2/2$ and $\theta_1=d\gamma+A$, 
we recover the resolved metric given above.

\section{Conclusion and Outlook}

In this paper, we have constructed symplectic potentials for a 
large family of metrics due to Apostolov, Calderbank and 
Gauduchon. We carry out a global analysis of these metrics, 
largely focusing on non-compact six-dimensional examples, by 
relating the symplectic potential to the canonical one due to 
Guillemin. We then systematically worked out the situations where 
we recover D-terms associated with known manifolds such as cones 
over $L^{pqr}$ and $Y^{pq}$ manifolds. We find among these 
metrics, an infinite family of partially resolved metrics for 
cones over $Y^{pq}$ for non-zero blow-up parameters. 
Interestingly, we also recover the resolved conifold (and its 
orbifold) and the resolution of $\BC^3/\BZ_3$ among the ACG 
metrics. Thus, all known examples of resolved metrics appear in 
this classification.

The $m=3$ orthotoric metrics seem the natural place to look for 
metrics corresponding to partial resolutions of $L^{pqr}$. In 
specific examples, we have found that there are no such solutions 
even though the blown down metric is recovered in a limit. 
Nevertheless, we feel that our analysis in this particular 
situation is incomplete and we hope to report on this in the 
future.

The paper has largely dealt with symplectic coordinates. One may 
wish to know if this is always a good approach. As a test case, 
we have attempted to work out the symplectic potential for 
resolutions of $\BC^2/\BZ_N$ using the symplectic quotient rather 
than the hyper-K\"ahler quotient that is natural in this setting. 
The symplectic method works only for $N=2$ but does \textit{not} work for 
$N>2$\cite{Aswinthesis}. However, it is known that a partial 
Legendre transform of the symplectic potential can be exactly 
determined in these examples\cite{Hitchin:1986ea} and an explicit 
map to the Gibbons-Hawking metrics worked out. In carrying out the
inverse Legendre transform to recover the symplectic potential, one
needs to find the roots of polynomials of degree greater than four
to come up with a closed-form expression for the symplectic potential.
Since no formulae exist for roots of polynomials with degree $>4$, one
does not obtain an algebraic expression the symplectic potential.

Our results clearly have implications in the context of the 
AdS-CFT correspondence. For instance, it is known that 
resolutions associated with two-cycles and four-cycles lead to 
different kinds of corrections to the radial part of the metric, 
i.e., $g_{rr}$\cite{Klebanov:1999tb}. These metrics provide an 
arena where this can be verified. The Abreu function that we have 
obtained in this paper may be used to verify the prediction of 
Martelli, Sparks and Yau on its behaviour\cite{Martelli:2006yb}.
Finally, the gravity dual of the (marginal) Leigh-Strassler
deformations of $\mathcal{N}=4$ supersymmetric Yang-Mills theory is
not yet known. The gravity dual is expected to have a $U(1)$ isometry
implying that it may arise from a $m=3$, $l=1$ ACG metric whose four-dimensional
base is a \textit{non-toric} K\"ahler-Einstein manifold. The CFT implies that the four-dimensional manifold must arise as a two-parameter deformation of  $\BC\BP^2$.

\noindent\textbf{Acknowledgments} AKB thanks 
the Department of Aerospace Engineering, IIT Madras and in particular,
Prof. Job Kurian and Prof. P.~Sriram for
encouragement and support. CNG thanks the hospitality of 
the theory group at IIT Madras and in particular, Prasanta K. 
Tripathy, for hosting a visit to IIT Madras during which the 
paper was completed.

\end{document}